\documentclass[]{iopart}

\usepackage{graphicx}
\usepackage{color}
\usepackage{amssymb}

\def\Xint#1{\mathchoice
   {\XXint\displaystyle\textstyle{#1}}%
   {\XXint\textstyle\scriptstyle{#1}}%
   {\XXint\scriptstyle\scriptscriptstyle{#1}}%
   {\XXint\scriptscriptstyle\scriptscriptstyle{#1}}%
   \!\int}
\def\XXint#1#2#3{{\setbox0=\hbox{$#1{#2#3}{\int}$}
     \vcenter{\hbox{$#2#3$}}\kern-.5\wd0}}

\def\dashint{\Xint-}

\newcommand{\fig}[1]{Figure (\ref{#1})}

\begin{document}

\title{Stochastic Loewner evolution driven by L\'evy processes}

\author{I Rushkin, P Oikonomou, L P Kadanoff and I A Gruzberg}
\address{The James Franck Institute, The University of Chicago,
5640 S. Ellis Avenue, Chicago, IL 60637}

\date{December 21, 2005}
\begin{abstract}

Standard stochastic Loewner evolution (SLE) is driven by a
continuous Brownian motion, which then produces a continuous
fractal trace. If jumps are added to the driving function, the
trace branches. We consider a generalized SLE driven by a
superposition of a Brownian motion and a stable L\'evy process.
The situation is defined by the usual SLE parameter, $\kappa$, as
well as $\alpha$ which defines the shape of the stable L\'evy
distribution. The resulting behavior is characterized by two
descriptors: $p$, the probability that the trace self-intersects,
and $\tilde{p}$, the probability that it will approach arbitrarily
close to doing so. Using Dynkin's formula, these descriptors are
shown to change qualitatively and singularly at critical values of
$\kappa$ and $\alpha$. It is reasonable to call such changes
``phase transitions''.  These transitions occur as $\kappa$ passes
through four (a well-known result)  and as $\alpha$ passes through
one (a new result). Numerical simulations are then used to explore
the associated touching and near-touching events.

\end{abstract}

\section{Introduction}

The scaling limit of many two-dimensional critical lattice models
and growth models encountered in statistical physics may be
described in terms of fractals. The boundaries of the
Fortuin-Kastelyn clusters in critical $q$-state Potts models and
critical percolation models, for example, in this limit are known
to be fractal, conformally invariant curves
\cite{nienhuis,duplantier}. Stochastic Loewner evolution
\cite{schrammorig,schramm} (also described as Schramm-Loewner
evolution and abbreviated as SLE) is a rigorous mathematical tool
for producing and studying stochastic conformally invariant curves
in the plane (see Refs. \cite{lawler,werner,kager,cardy} for
review). It was conjectured, therefore, that SLE is a description
of such statistical systems. In several special cases this
conjecture has been rigorously proven.

SLE is based on the Loewner equation taken from complex analysis
\cite{ahlfors}. This equation contains a driving (or forcing)
function, $\xi(t)$, which determines all the properties of SLE. A
few reasonable requirements constrain the choice of this function
to a scaled Brownian motion, which leaves only one free parameter.
One of these requirements (that the forcing function is
continuous) ensures that the curves produced by SLE do not exhibit
branching, thus leaving many systems, such as branching polymers,
out of the picture. In this paper, we generalize SLE by dropping
the demand that the forcing stochastic function be continuous, but
keeping the requirement of stationary and
statistically-independent increments. This leads to a much broader
class of forcing processes including, in particular, the so called
L\'evy processes \cite{apple,sato,ta,metz}. This generalization
might be a useful description of tree-like stochastic growth.

Standard SLE has been studied in various geometries under the
names of chordal, radial and other kinds of SLE's. In this paper
we restrict ourselves to the chordal situation (growth in the
upper half plane). We believe that our generalized SLE in a radial
geometry may be relevant for description of the diffusion-limited
aggregation, and plan to study it in a separate publication.

In the next section, we define and describe the problem, in terms
of the parameters which define the forcing. The forcing includes
both a scaled Brownian motion and a stable L\'evy process. The
parameters are $\kappa$, which sets the normalization of the
Brownian term, $c$ which sets the normalization of the L\'evy
term, and $\alpha$ which determines its shape. This qualitative
section ends with numerical simulations of the traces, the
geometrical structures generated by the SLE.

The final section of the paper gives the analysis of the
short-distance properties of the traces showing both analytically
and numerically that the traces have qualitative change in
behavior as $\kappa$ and $\alpha$ each pass though critical
values, respectively at four and one. The transition at $\kappa =
4$ is quite analogous to a well known transition in the standard
SLE \cite{schramm}. The latter phase transition, at $\alpha = 1$,
is entirely new. This paper focuses upon the short-distance
manifestations of this phase transitions in the chordal geometry.
The accompanying large-scale features (which may be different for
different geometries) will be the subject of a subsequent paper.

We point out that most of the statements about the behavior of
objects related to SLE and its generalizations are probabilistic,
and typically appended by qualifiers ``almost surely'' or ``with
probability one'' in mathematical literature. We will not use
these qualifiers explicitly assuming that they should be applied
wherever it is necessary.

Some technical details of our calculations are given in the
Appendix.

\section{Description of problem}

\subsection{Loewner evolution}

The so called chordal Loewner evolution is described by a
differential equation obeyed by a function $g(z,t)$. At each value
of the ``time'' $t$, this function is a conformal map from a
subset of the upper half of the complex $z$-plane (which we refer
to as the {\it physical plane}) to the entire half plane $H$ ({\it
mathematical plane}). Specifically we write
\begin{equation}
\label{11}
\partial_t g(z,t) = \frac{2}{g(z,t) - \xi(t)}, \qquad
g(z,0) = z.
\end{equation}
Here $\xi(t)$ is the forcing, mentioned above. Alternatively, the
function $h(z,t) = g(z,t) - \xi(t)$ obeys
\begin{equation}
\label{11h}
\partial_th(z,t) = \frac{2}{h(z,t)} -
\partial_t\xi(t), \qquad h(z,0) = z,
\end{equation}
assuming that $\xi$ vanishes at $t=0$.

\begin{figure}[t]
\centering
\includegraphics[scale=0.3]{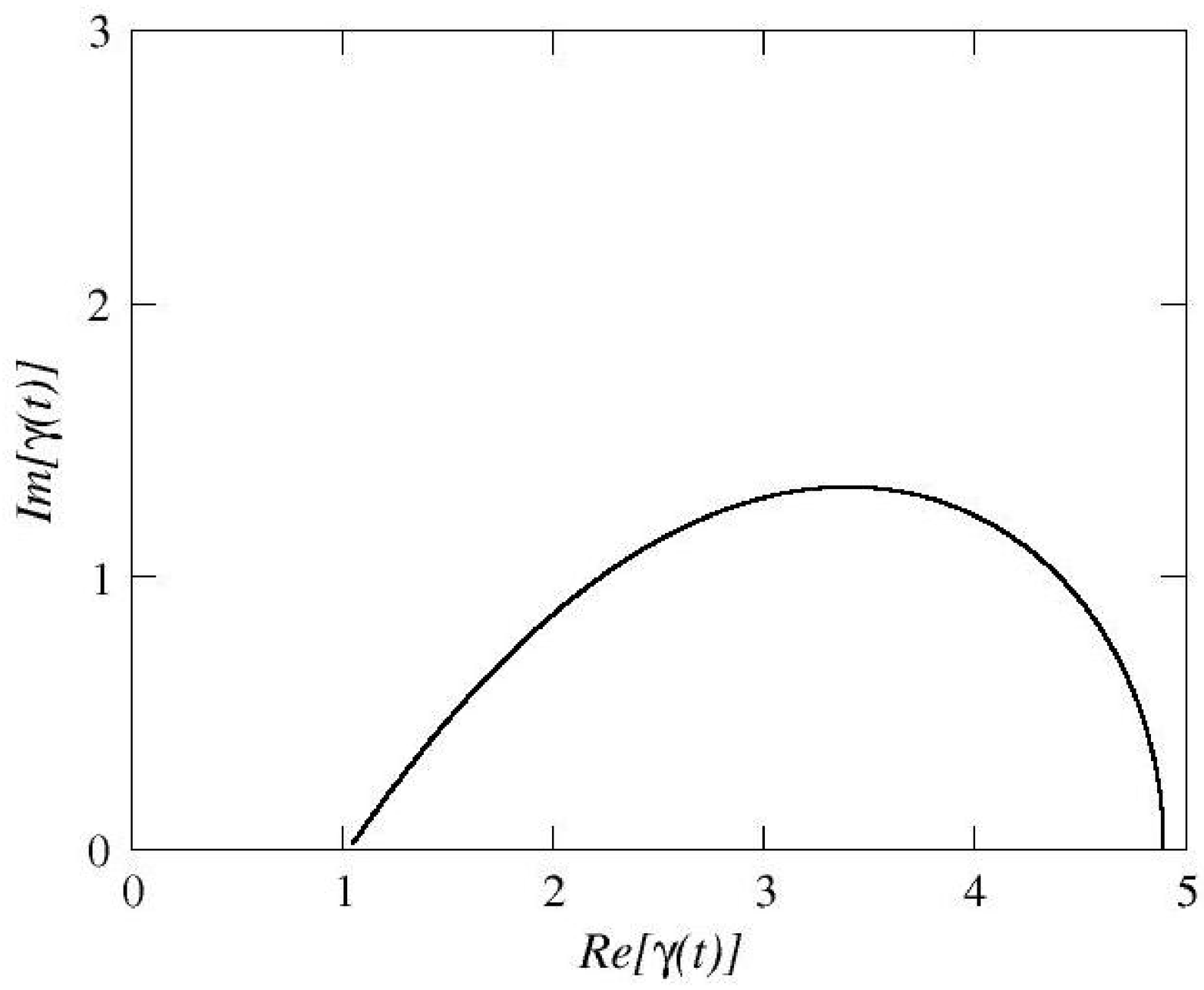}
\includegraphics[scale=0.3]{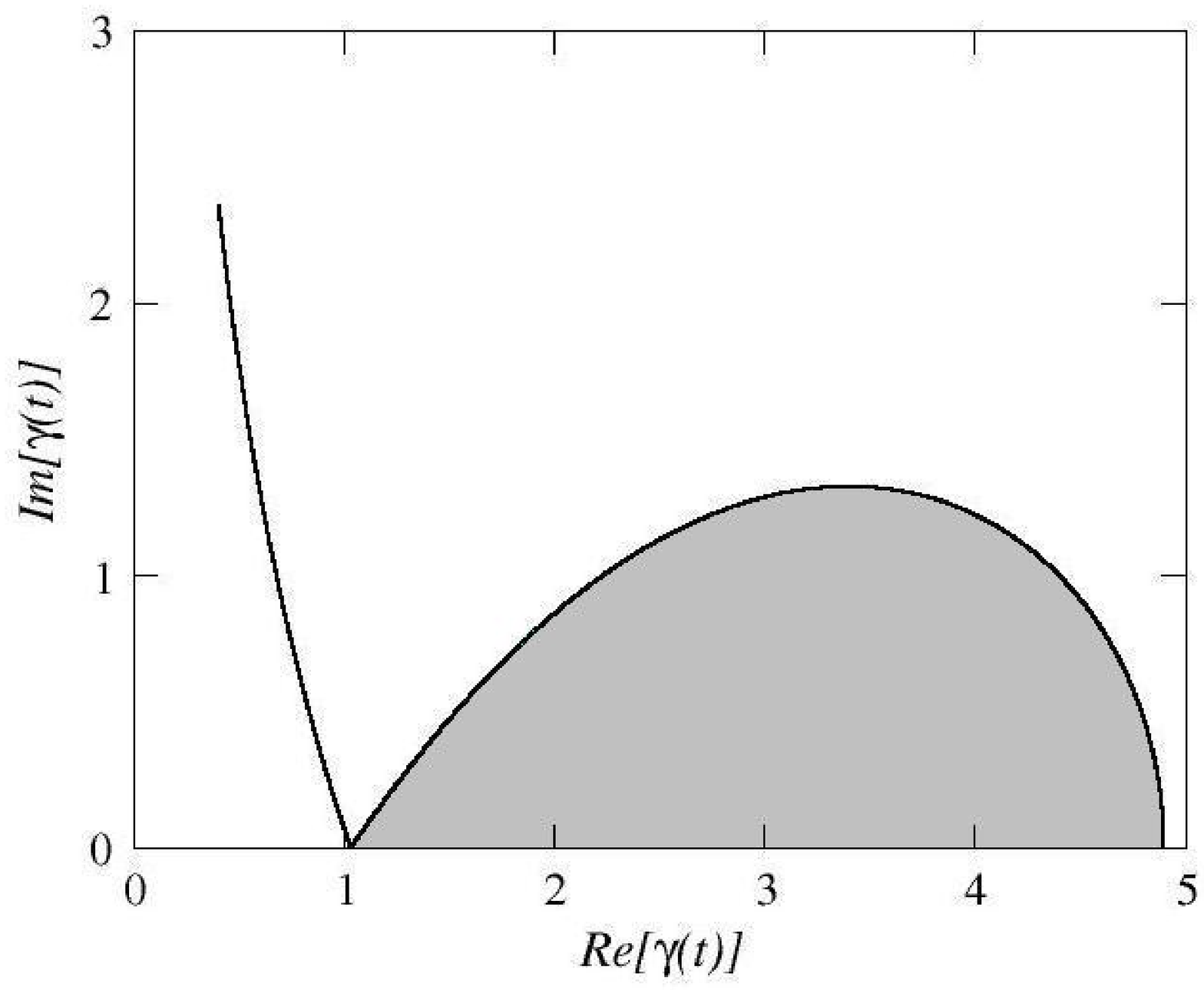}
\caption{The trace and the hull for a touching event. Here
$\xi(t)=2\sqrt{6(1 - t)}$ for $t \in (0,1)$ and zero elsewhere
\cite{exactLPK}. Left shows the situation just before touching ($t
\to 1^{-}$); right shows the situation just after ($t > 1$). The
trace is the thick dark line. The hull consists of that line plus
the grey area. That area is added to the hull at $t = 1$. Note
that there is a continuum of points added to the hull at the time
of touching, but only one of these is on the hull's boundary and
is the point $\gamma(1)$ of the trace.} \label{fig1}
\end{figure}

For any point $z \in H$ these equations are valid up to the
swallowing time $\tau(z)$, which is defined as the first time when
$g(z,t) = \xi(t)$ or $h(z,t) = 0$ so that the right hand sides of
Eqs. (\ref{11}) and (\ref{11h}) become singular. The {\em hull} of
the Loewner evolution is defined as the set of all points which
become singular (are swallowed) in this way up to time $t$. At any
given time, the $z$'s that newly enter the hull may represent a
single point or a more complicated set, sometimes including a
subset $H$ with a finite area. See Figure \ref{fig1}.

Another geometrical object considered in the SLE literature is the
growing tip $\gamma(t)$ of the hull. The point $z=\gamma(t)$ is
picked from among the (perhaps many) points which newly enter the
hull at time $t$. Among these, this point is unique for having the
property of being on the boundary of the hull at the time $t$, see
Figure \ref{fig1}. More formally, this point is defined by
\cite{schramm}
\begin{equation}
\gamma(t) = \lim_{z \to 0} h^{-1}(z,t),
\end{equation}
where the limit is taken within $H$. The union of the points
$\gamma(t)$ up to a time $T$ is the trace of the evolution up to
$T$.

The choice of the process $\xi(t)$ defines the properties of the
hull and the trace. In particular, if $\xi(t)$ has a discontinuity
(jumps), branching or discontinuity occurs in the trace and the
hull may become disconnected. For a sufficiently smooth $\xi(t)$
(a precise condition involves H\"older continuity, see Refs.
\cite{Marshall-Rohde,exactLPK,lind}) the trace is a curve
parametrized by $t$. Moreover, if the trace is a simple curve,
that is a curve without double points and without touchings of the
boundary, it then coincides with the hull.

\subsection{The forcing}

Standard SLE \cite{schrammorig} is the Loewner evolution with
scaled Brownian forcing
\begin{equation}
\xi(t)=\sqrt{\kappa} B(t), \label{SLEU}
\end{equation}
where $B(t)$ is a standard Brownian motion. In physical terms this
means that the ``time derivative'' $\dot \xi(t)$ is a white noise
with intensity $\kappa$: $\langle \dot \xi(t) \dot \xi(t') \rangle
= \kappa \delta(t - t')$. The increments of $\xi(t)$ in distinct
time intervals are stationary and statistically independent. These
properties of $\xi(t)$ ensure that the process $g(z,t)$ is
conformally invariant in the sense that the time evolution is
consistent with composition of conformal maps. In particular, the
ensemble of $g(t + s)$ can be obtained by functional composition
of $g(t)$ with $g(s)$, each drawn from the correct ensemble.

If we want to generalize Eq. (\ref{SLEU}) but maintain the
functional composition property, we may only discard the
requirement of continuity of the forcing $\xi(t)$ but keep its
increments to be stationary and statistically independent. This
leads to a broad class of stochastic processes, including the so
called L\'evy processes \cite{apple,sato,ta}, also called L\'evy
flights in physics literature \cite{metz}. The simplest of these
are the so called stable L\'evy processes $L_\alpha(t)$
characterized by a single real parameter $\alpha$ in the range $0
< \alpha < 2$. The process $L_\alpha(t)$ is composed of a
succession of jumps of all sizes. In a time interval $dt$,
$L_\alpha(t)$ jumps by an amount between $x$ and $x + dx$ with the
probability proportional to
\begin{equation}
dt \, dx \, \frac{1}{|x|^{1 + \alpha}}.
\end{equation}
More precisely, the probability distribution function of
$L_\alpha(t)$ is given by the Fourier transform
\begin{equation}
P(x,t) = \int_{-\infty}^\infty e^{-ikx} e^{-t |k|^\alpha}
\frac{dk}{2\pi}.
\end{equation}

The case of $\alpha = 2$, formally corresponding to a Brownian
motion, is, in fact, quite different from $\alpha < 2$. We find it
interesting to combine a Brownian motion with a stable L\'evy
process as the input to our generalized version of the SLE. Thus
we choose as our forcing the function
\begin{equation}
\xi(t)=\sqrt{\kappa} B(t) + c^{1/\alpha} L_\alpha(t).
\label{SLEUL}
\end{equation}
Every time a discontinuity of the driving force occurs the trace
develops a branching point. With L\'evy flights, therefore, there
are infinitely many branching points on all size-scales.

The stable L\'evy processes are self-similar \cite{apple}. For any
positive number $b$ the two processes, $L_\alpha(t)$ and
$b^{1/\alpha}L_\alpha(t/b)$, are ``equal in distribution'', that
is, statistically identical, which we denote as
\begin{equation}
\label{self-similarity} L_\alpha(bt) \stackrel{d}{=} b^{1/\alpha}
L_\alpha(t).
\end{equation}
Thus, a typical trajectory of a L\'evy flight looks like many
clusters separated by long jumps. Each cluster, if zoomed into,
again looks like many clusters separated by long jumps, and so on
\cite{metz}.

\subsection{Scales for SLE with L\'evy flights}

We choose the driving force $\xi(t)$ in Loewner evolution to be of
the form (\ref{SLEUL}):
\begin{equation}
\label{12}
\partial_t g(z,t) = \frac{2}{g(z,t) - \sqrt{\kappa}B(t) -
c^{1/\alpha}L_{\alpha}(t)}, \qquad g(z,0) = z.
\end{equation}
If $c = 0$ this process is scale-invariant \cite{kager}. The run
of the Loewner evolution up to a time $bt$ is statistically the
same as the run up to a time $t$, rescaled by a factor of
$\sqrt{b}$. The properties of the growing hull do not change with
time except for general size rescaling. In this Brownian driving
force is exceptional.

The addition of L\'evy flights changes the situation: time becomes
significant. Consider a time rescaling $t \to  bt$, where $b$ is a
positive number. We construct  the rescaled conformal map
\begin{equation}
g_b(z,t) = b^{-\frac{1}{2}}g(b^{\frac{1}{2}}z, bt).
\end{equation}
It satisfies the initial condition $g_b(z,0) = z,$ and using the
self-similarity property (\ref{self-similarity}) we observe that
$g_b$ obeys an equation of exactly the same form as Eq. (\ref{12})
except that the value of $c$ is changed to
\begin{equation}
c_b = cb^{(2 - \alpha)/2}. \label{resc}
\end{equation}
The strength of the L\'evy process flows under time-rescaling,
making SLE not scale-invariant. (Many calculations in the usual
SLE with pure Brownian motion are based on scale-invariance
\cite{schramm,werner,kager,beffara} so that they cannot
immediately be extended to the present case.)

This situation can be phrased in the language of the
renormalization group (RG). The critical behavior of a system
described by an SLE with purely Brownian forcing is determined by
a fixed point of an RG. L\'evy flights make the system non-scale
invariant, and their addition to the forcing can be thought of as
a perturbation that is relevant at the fixed point. We do not know
at present what such perturbations might be in terms of
microscopic models of critical systems described by Brownian SLE.
One possible candidate is a long range interaction of some kind.

The flow of $c$ has a fixed point at zero, when we recover the
result that SLE with pure Brownian motion is scale-invariant. For
$c>0,$ the ratio of $\kappa$ and the effective $c$ changes with
time. At time $t=1$ it is just $ c_1  =  c.$ At time $t$ the
effective ratio of the two terms in the forcing is
\begin{equation}
\frac{c^{1/\alpha}_{\rm eff}}{\kappa^{1/2}} =
\frac{c^{1/\alpha}}{\kappa^{1/2}}t^{\frac{1}{\alpha} -
\frac{1}{2}}.
\end{equation}
Therefore, crossover behavior can be observed at a time
\begin{equation}
t^* = \Big(\frac{\kappa^\alpha}{c^2}\Big)^{1/(2 - \alpha)}.
\end{equation}
At small times $t\ll t^*$ the evolution is dominated by the
Brownian motion, whereas at large times $t\gg t^*$ it is dominated
by L\'evy flights.  (If $\kappa$ were very large or very small one
would have to consider the independent effect of the three terms
in the denominator of Eq. (\ref{12}), but for $\kappa$ of order
one this is the only crossover that need be considered.) There is
a similar crossover in space at a spatial scale $x^* =
\sqrt{t^*}$, with the Brownian motion dominating the small scale
events.

\subsection{Numerical representation of SLE}

We approximate the integration of the SLE by introducing a
discretized version, where each realization of $\xi(t)$ is
replaced by a piecewise constant function with jumps appropriately
distributed. Let $X_i$, $i = 1, 2, \ldots$, be a sequence of real
stochastic variables. We define a realization of the discretized
driving force to be constant in the interval $((j - 1)\tau,\
j\tau]$ (where $\tau$ defines the mesh on the time variable) with
$\xi(0)=0$ and
\begin{equation}
\label{eq:FORCING} \xi(t) \to  \xi_j = \sum_{i=1}^j X_i \qquad
\textrm{for } \quad (j - 1)\tau < t \leqslant j\tau.
\end{equation}
For this forcing we can exactly solve for the trace. Within each
time interval $(t_{j - 1},\ t_j)$ we can solve the SLE forward:
\begin{equation}
\label{eq:SOL} g(z, j\tau)=\sqrt{[g(z, (j - 1)\tau) - \xi_j]^2 +
4\tau} + \xi_j,
\end{equation}
or backward:
\begin{equation}
\label{eq:BCK-SOL} g(z, (j - 1)\tau)=\sqrt{[g(z, j\tau) - \xi_j]^2
- 4\tau} + \xi_j.
\end{equation}
Equations (\ref{eq:SOL}), (\ref{eq:BCK-SOL}) define an
infinitesimal conformal map and its inverse for each time interval
$j$:
\begin{eqnarray}
w_j(z) = \sqrt{(z - \xi_j)^2 + 4\tau} + \xi_j, \\
f_j(z) = w_j^{-1}(z) = \sqrt{(z - \xi_j)^2 - 4\tau} + \xi_j.
\label{eq:MAP}
\end{eqnarray}
The trace can then be calculated numerically as an iteration
process of infinitesimal conformal mappings starting from the
condition $g(\gamma(t),t) = \xi(t)$ as follows
\cite{hastings,singularLPK}:
\begin{equation}
\label{eq:ITERATE-MAP} \gamma_j = \gamma(j\tau) = f_1 \circ f_2
\circ \ldots f_{j-1} \circ \ f_j (\xi_j).
\end{equation}
The value of $\xi_j$ is determined by the variables $X_i$ which
are randomly drawn from the appropriate distribution according to
the properties of the driving force $\xi(t)$.

\begin{figure}
\centering
\includegraphics[scale=0.55]{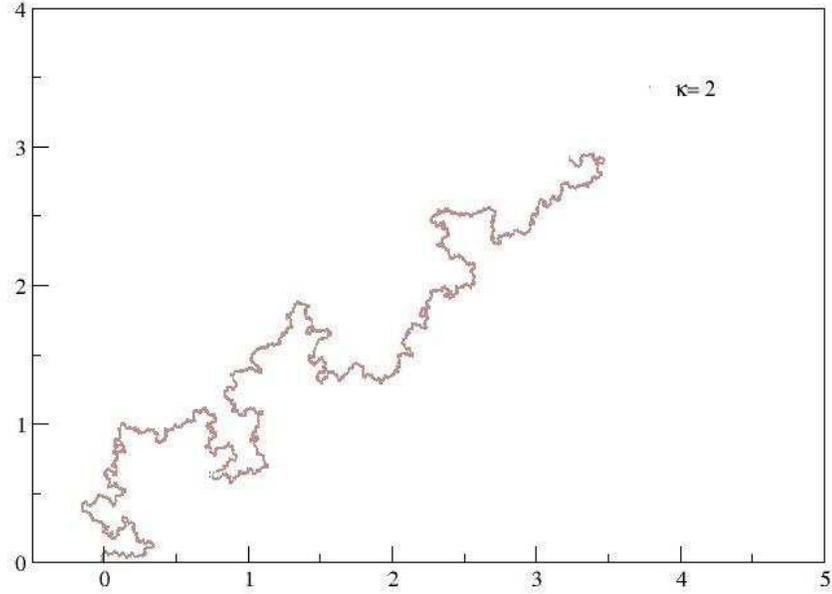}
\caption{SLE trace for Brownian forcing $\kappa=2$, $300000$
steps, $\tau=10^{-5}$, $T=3$} \label{pick2}
\end{figure}

\begin{figure}
\centering
\includegraphics[scale=0.55]{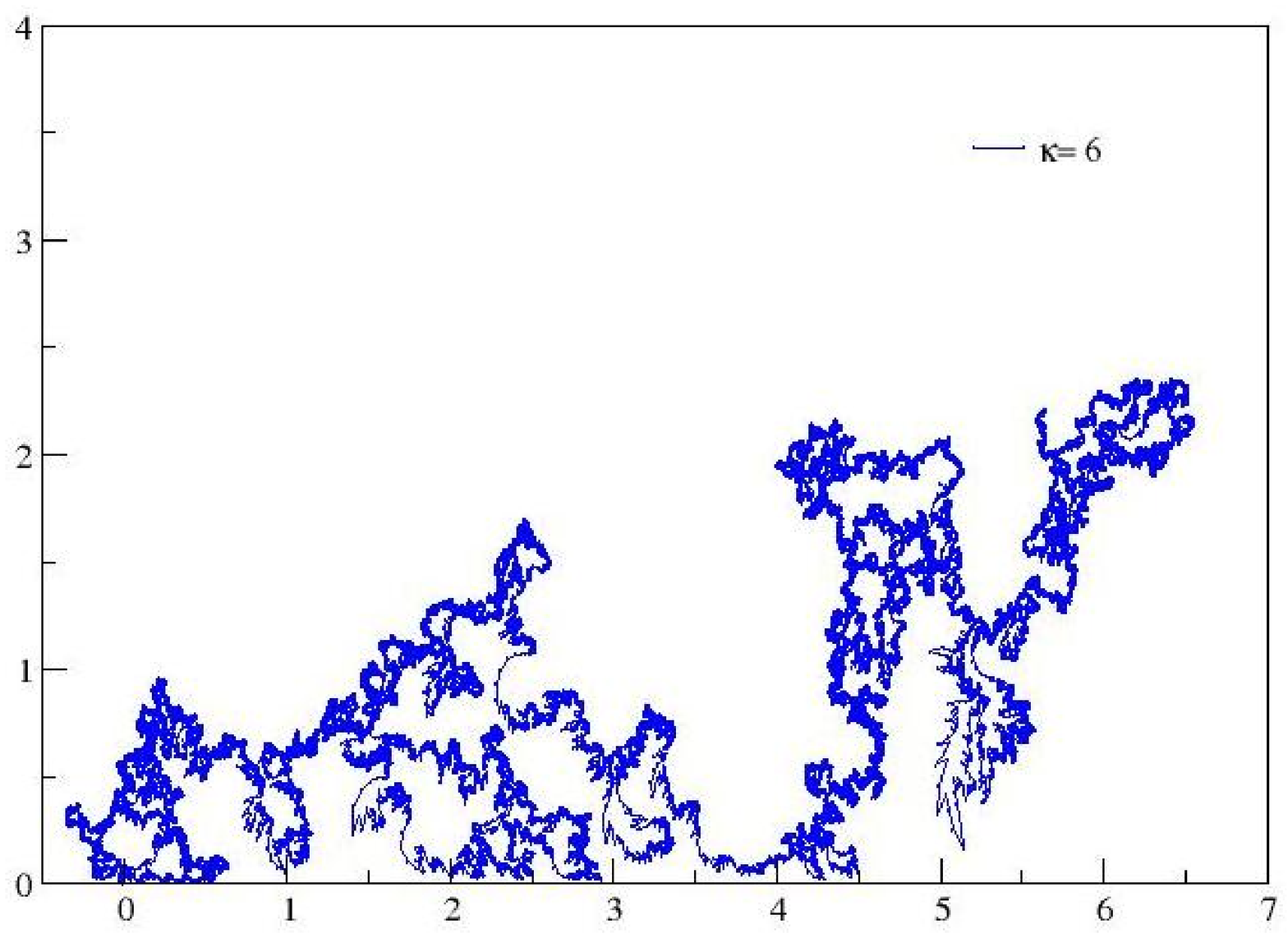}
\caption{SLE trace for Brownian forcing $\kappa=6$, $300000$
steps, $\tau=10^{-5}$, $T=3$} \label{pick6}
\end{figure}

\begin{figure}
\centering
\includegraphics[scale=0.55]{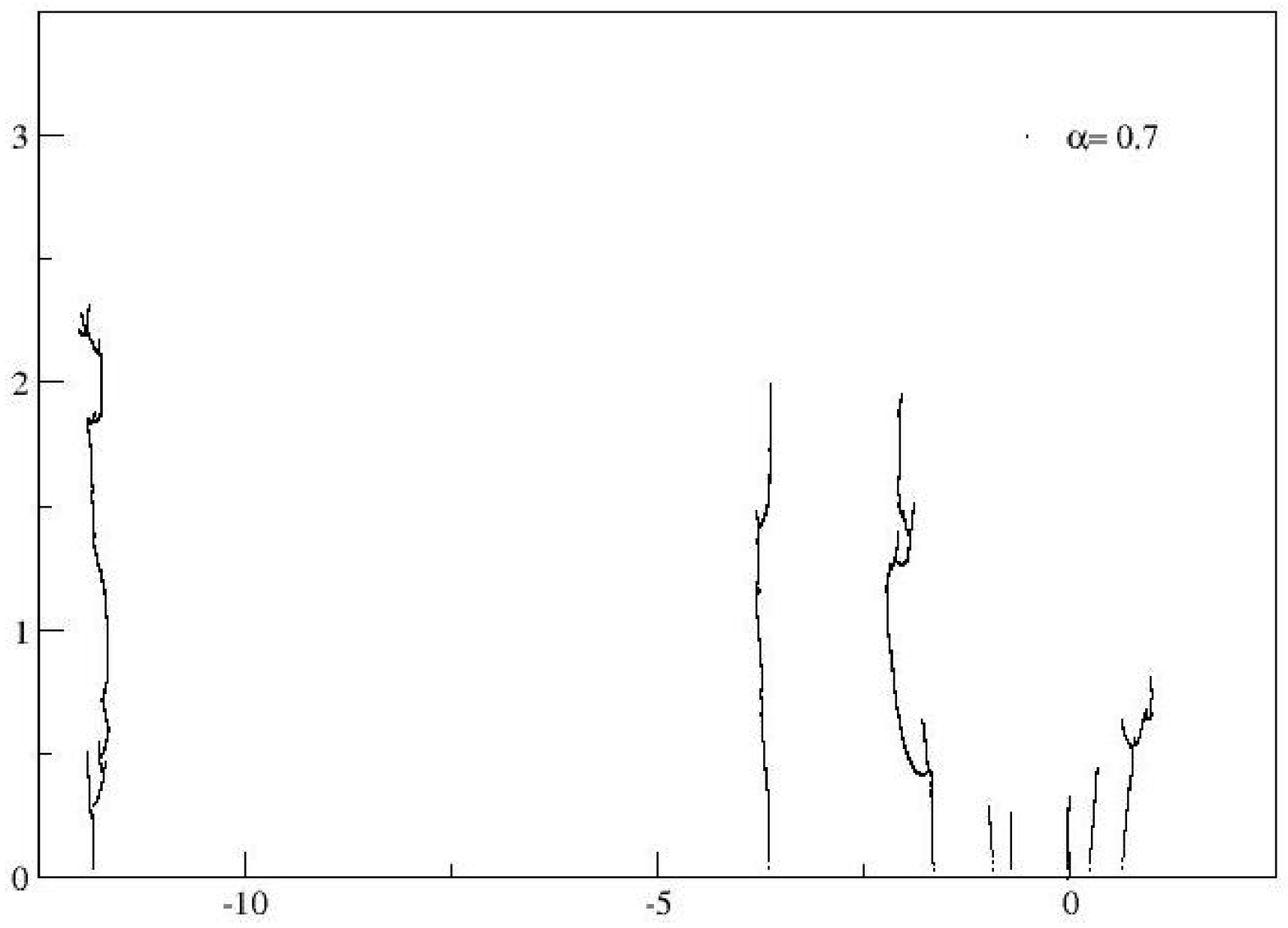}
\caption{SLE trace for L\'evy distributed forcing $\alpha=0.7$,
$c=4$, $300000$ steps, $\tau=10^{-5}$, $T=3$} \label{picm07}
\end{figure}

\begin{figure}
\centering
\includegraphics[scale=0.55]{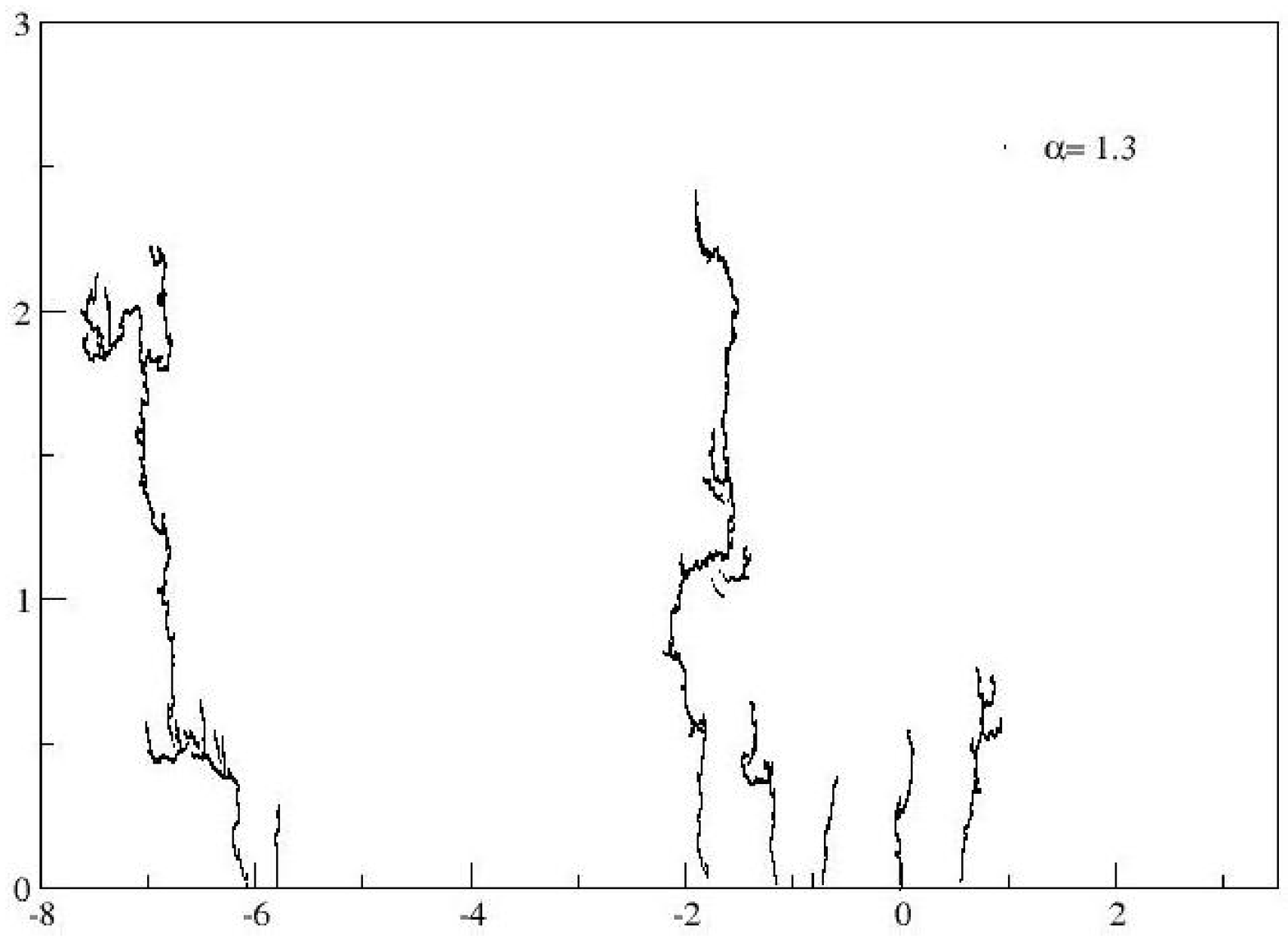}
\caption{SLE trace for L\'evy distributed forcing $\alpha=1.3$,
$c=4$, $300000$ steps, $\tau=10^{-5}$, $T=3$} \label{picm13}
\end{figure}

\begin{figure}
\centering
\includegraphics[scale=0.55]{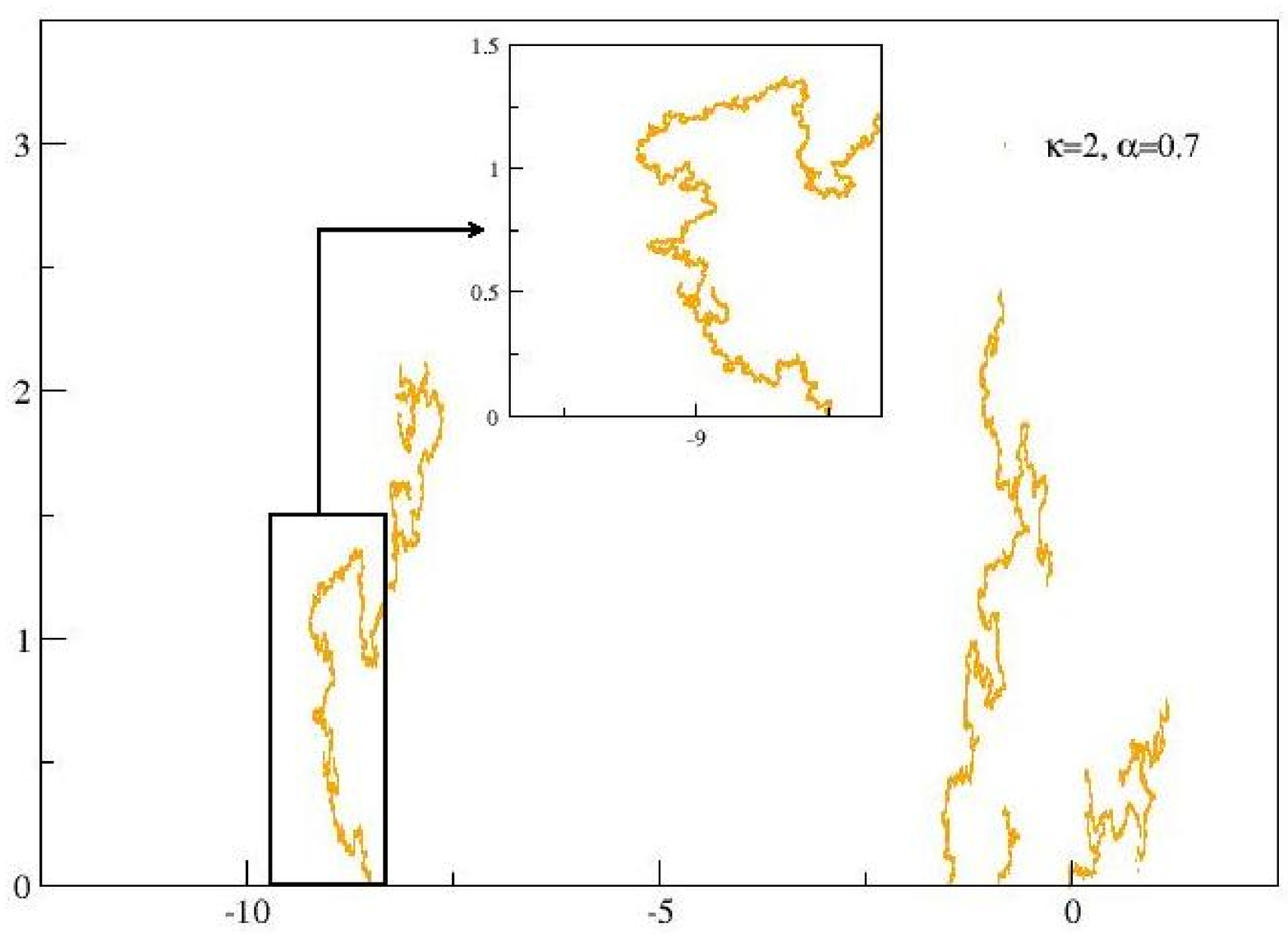}
\caption{SLE trace with both L\'evy flights and Brownian forcing
$\alpha=0.7$, $c=4$, $\kappa=2$, $300000$ steps, $\tau=10^{-5}$,
$T=3$, $t^*=0.17$} \label{pick2m07}
\end{figure}

\begin{figure}
\centering
\includegraphics[scale=0.55]{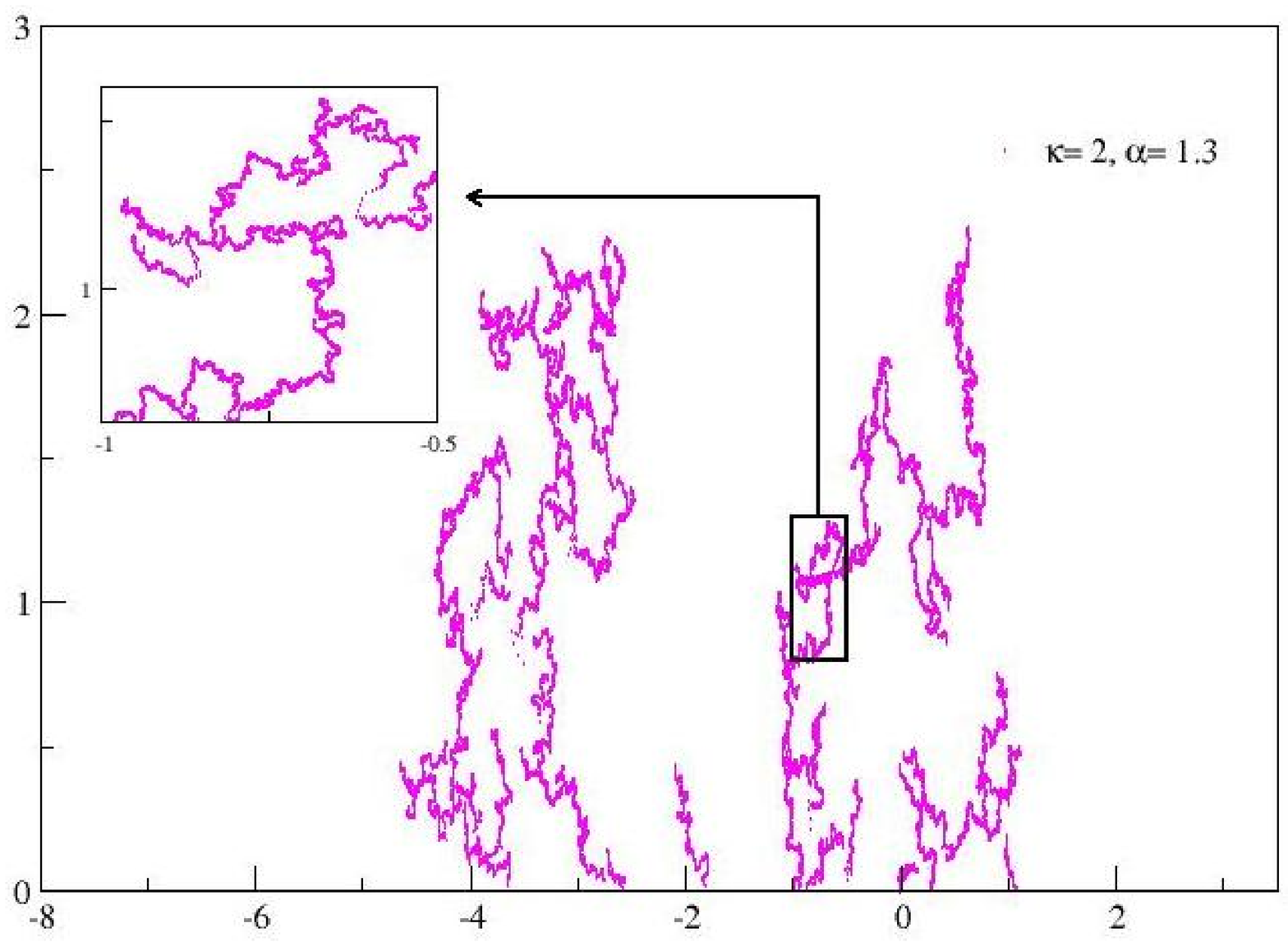}
\caption{SLE trace with both L\'evy flights and Brownian forcing
$\alpha=1.3$, $c=4$, $\kappa=2$, $300000$ steps, $\tau=10^{-5}$,
$T=3$, $t^*=0.31$} \label{pick2m13}
\end{figure}

\begin{figure}
\centering
\includegraphics[scale=0.55]{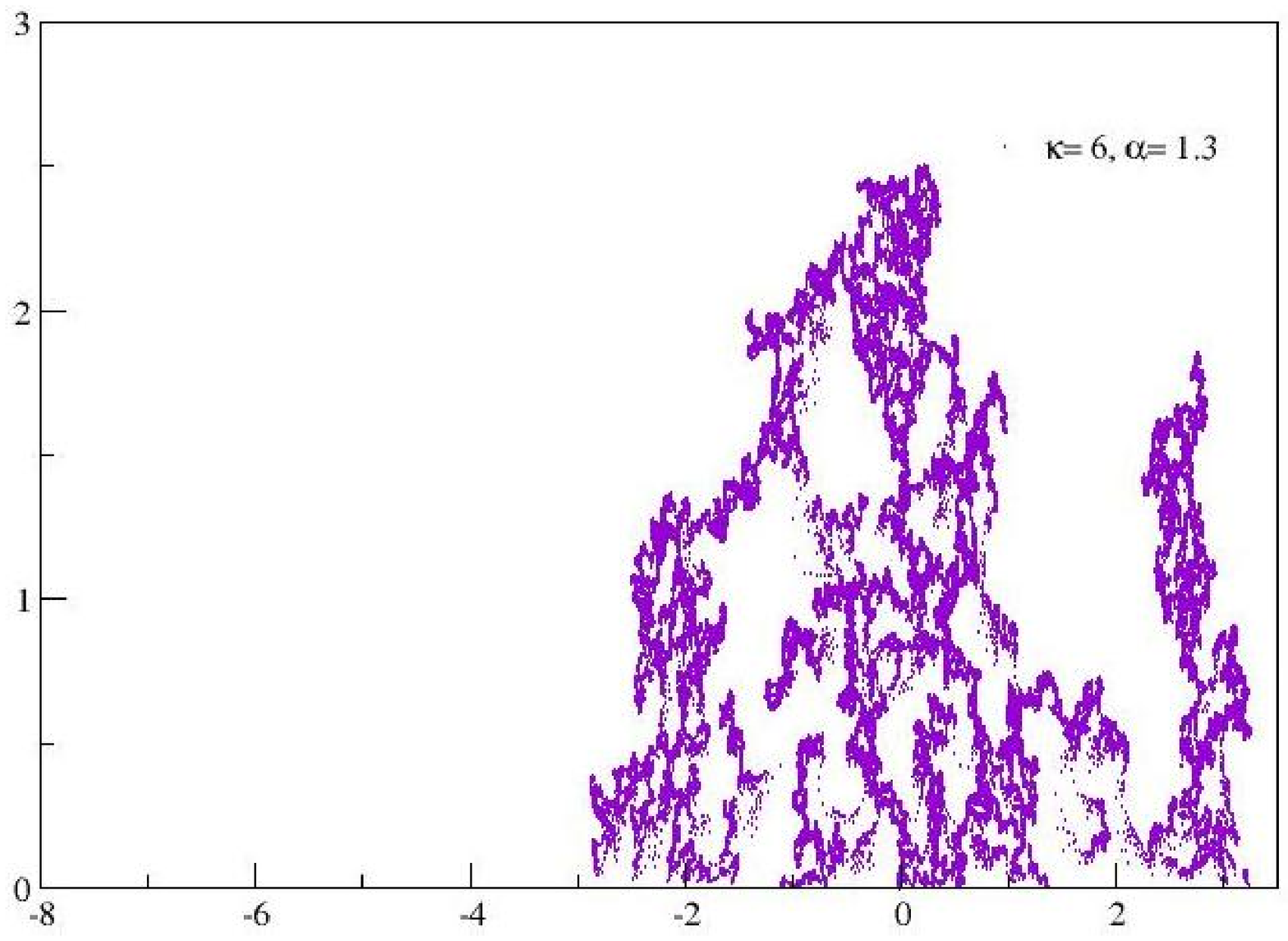}
\caption{SLE trace with both L\'evy flights and Brownian forcing
$\alpha=1.3$, $c=4$, $\kappa=6$, $300000$ steps, $\tau=10^{-5}$,
$T=3$, $t^*=0.07$} \label{pick6m13}
\end{figure}

\begin{figure}
\centering
\includegraphics[scale=0.55]{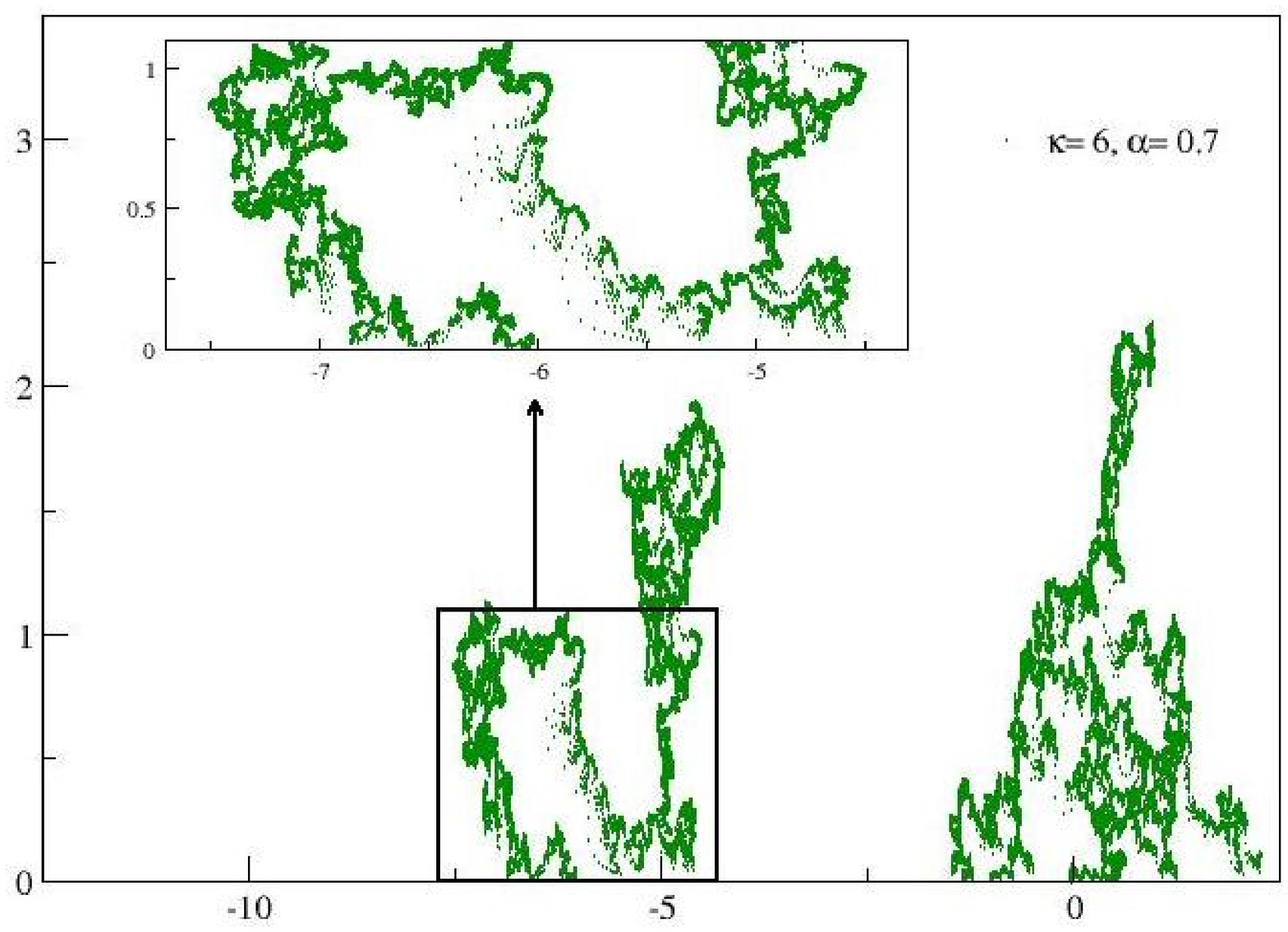}
\caption{SLE trace with both L\'evy flights and Brownian forcing
$\alpha=0.7$, $c=4$, $\kappa=6$, $300000$ steps, $\tau=10^{-5}$,
$T=3$, $t^*=0.53$} \label{pick6m07}
\end{figure}

In Figures \ref{pick2}--\ref{pick6m07} we show typical SLE traces
for different types of driving force up to a time $T$. For Figures
\ref{pick2}--\ref{pick6} we have set $c=0$ and used increments
drawn from Gaussian random variables in order to show SLE traces
for the standard case of Brownian noise.   The noise realization
for these two pictures is the same, multiplied by an appropriate
$\kappa$. As is well known \cite{schramm},\cite{kager}, the trace
does not touch itself for $\kappa \leqslant 4$, but does so for
$\kappa>4$. The figures show these contrasting behaviors.

For Figures \ref{picm07}--\ref{picm13} we approximate SLE with
L\'evy flights using the method in \cite{chambers}. As expected,
these figures show branching structures produced by the
discontinuous jumps.  The figures suggest that a forcing
consisting solely of discontinuous jumps does not produce any
self-intersections.

Figures \ref{pick2m07}--\ref{pick6m07} show the combined effects
of jumps and Brownian motion. Exactly the same noise realizations
have been used as in the previous four figures. The different
sources of change in $X$ (Brownian and L\'evy) have simply been
added to one another.

Comparison between Figures \ref{pick2m07}--\ref{pick6m07} and
\ref{pick2}--\ref{picm13} exemplifies the crossover behavior
discussed in the previous section. Take for example \fig{picm07}.
The trace consists of branches which look like the ones produced
by the Loewner equation with constant forcing. Whenever a jump
occurs, a new branch starts growing off some point on the trace or
the real axis. If now we add Brownian noise to the driving force
(Figures \ref{pick2m07}--\ref{pick6m07}) the trace at the largest
scales makes branches like in the case with only L\'evy flights.
However, if we zoom in (see insets) the trace appears similar to
the SLE trace with only Brownian forcing (Figures
\ref{pick2}--\ref{pick6}).

\section{Analysis of short-distance behavior}

\subsection{Phases of SLE}

It is well known \cite{schramm} that the standard Brownian SLE
process is qualitatively different in its behavior in the three
different intervals $0 < \kappa \leqslant 4$, $4 < \kappa < 8$,
and $8 \leqslant \kappa < \infty$.  In the first interval the SLE
trace forms a fractal that neither touches itself nor the real
line (see Figure \ref{pick2}). In the second interval the trace
touches itself and the real line (see Figure \ref{pick6}). For
large enough time it will touch down arbitrarily far from the
origin. Eventually it will surround and swallow up any point on
the real line no matter how far it is from the origin. In the
third, all points on the real line are surrounded sequentially,
but also the trace fills in all the surrounded areas. This kind of
behavior is not discussed here. The different behaviors for
different values of $\kappa$ are described as different phases of
SLE, basically because the behavior in each interval is
qualitatively different from that in the others. The difference is
topological in character. It can be seen in both the small- and
the large-scale features of the trace. Here we shall discuss
phases of the model containing both Brownian and L\'evy forcing,
but concentrate mostly on small-scale features of touching and
near-touching.

Notice that the events when the trace touches the real line and
the previously existing parts of itself, are equivalent in the
statistical sense, which follows from the stationary increments of
the forcing and the composition property of conformal maps. Since
the driving force $\xi(t)$ is stationary and SLE is defined by a
first-order differential equation, we have the statistical
identity
\begin{equation}
h(h(z,s),t)\stackrel{d}{=}h(z, s + t).
\end{equation}
Consider the trace grown up to a time $s + t$. The part of the
trace that developed earlier (up to time $s$) can be mapped to the
real axis by $h(z,s)$. Then the event of the growing hull
swallowing (or coming arbitrarily close to swallowing) a point
somewhere on its previously grown boundary is statistically
equivalent to the event of the hull swallowing (or almost
swallowing) a point on the real axis, and the probabilities of
such events should be the same. Therefore, to determine the
probability of swallowing for points anywhere on the boundary of
the hull, it is enough to determine the probability of swallowing
only for the points on the real axis. This was initially argued
for purely Brownian SLE, and remains valid here.

\subsection{Using Dynkin's formula}

The main tool that we will use to determine phase transitions in
SLE is Dynkin's formula \cite{oksendal,encyclo}. It can be written
for a generic stochastic process (not necessarily L\'evy), but we
will restrict it to L\'evy processes and stochastic integrals of
them, that is, processes $X(t)$ defined by a Langevin-like
equation
\begin{equation}
\partial_t X_t = F(X(t))  + \partial_t(\textrm{L\'evy
process}),
\end{equation}
with any drift function $F$. Dynkin's formula gives the average
value of any function $f$ of the stochastic process $X(t)$ at an
exit time $T$:
\begin{equation}
\langle f(X(T))\rangle = f(x_0)  +  \Big\langle \! \int_0^T \!\!\!
Af(X(t))dt \Big\rangle.
\end{equation}
Here $x_0$ is the initial value of the process: $X(0) = x_0,$ and
$A$ is the generator of the process, that is, the operator acting
on the argument of the function $f$ and representing time
differentiation. By choosing $f(x)$ to be a zero mode of the
generator:
\begin{equation}
Af(x) =0, \label{zero mode}
\end{equation}
we get rid of the integral term. Note that we may solve (\ref{zero
mode}) with any boundary condition: any non-constant solution is
sufficient. In the Brownian case this equation is differential and
easy to solve. Thus, Dynkin's formula produces many results about
Brownian motion, such as recurrence, transience, etc.
\cite{oksendal}.

\subsection{Brownian forcing}

As an example of using Dynkin's formula we first derive the phase
transition at $\kappa = 4$ for purely Brownian SLE. This
calculation is usually done in literature with the aid of
martingales \cite{schramm}.

Let us set $c = 0$ in Eq. (\ref{SLEUL}). As a function of time,
$h(z,t)$ is a continuous stochastic process with the initial
condition $h(z,0) = z.$ The event in which $z$ is swallowed by the
hull at a time $T$ corresponds to the event $h(z,T) = 0$ on the
mathematical plane. We choose $z$ to be on the real axis in the
physical plane, then $h(z,t)$ is real.

Let us consider a real $x > 0$ for concreteness. On the
mathematical plane, we fix points $a$ and $b$ on the real axis so
that
\begin{equation}
0 < a < x < b < \infty.
\end{equation}
Let $T$ be the exit time from the interval $[a,b]$. Being
continuous, the process $h(x,t)$ can exit $[a,b]$ either through
$a$ (with probability $p_a$), or through $b$ (with probability $1
- p_a$). Clearly, $h(x,T)$ is either $a$, or $b$. Then Dynkin's
formula becomes
\begin{equation}
p_a f(a)  + (1 - p_a)f(b) = f(x),
\end{equation}
and allows us to find $p_a$, the probability for $h(x,t)$ to hit
$a$ before hitting $b$. Now if we take the limits $a  \to 0,\, b
\to \infty,$ it becomes the probability for $h(x,t)$ to hit $0$ in
a finite time, which is the probability for the point $x$ on the
physical plane to be swallowed by the hull:
\begin{equation}
\label{40} p = \lim_{b \to \infty} \lim_{a \to  0} \frac{f(x) -
f(b)}{f(a) - f(b)}.
\end{equation}
In general, the order of limits matters here. If it is reversed,
\begin{equation}
\label{50} \tilde{p} = \lim_{a \to 0}\lim_{b \to \infty}\frac{f(x)
- f(b)}{f(a) - f(b)}
\end{equation}
is the probability for $h(z,t)$ to come arbitrarily close to 0,
that is, for the hull to come arbitrarily close to the boundary.

The generator for the process $h(z,t)$ driven by the Brownian
motion (\ref{SLEU}) is \cite{oksendal}
\begin{equation}
A_B = \frac{2}{x}\partial_x  + \frac{\kappa}{2}\partial_x^2.
\label{A_B}
\end{equation}
A zero mode of this operator is
\begin{equation}
\label{f_B} f_B(x) = \left\{
\begin{array}{ll}
|x|^{1 - \frac{4}{\kappa}} & \textrm{for $\kappa\neq 4$}, \\
\log |x| & \textrm{for $\kappa = 4$}.
\end{array}
\right.
\end{equation}
Substituting this into Eqs. (\ref{40}, \ref{50})  we find that the
answer is independent of $z$, thus the probability for the hull to
touch the boundary
\begin{equation} p_{c=0} = \left\{
\begin{array}{ll}
0 & \textrm{for $\kappa \leqslant 4$}, \\
1 & \textrm{for $\kappa > 4$}.
\end{array}
\right.
\end{equation}
For $\kappa=4$ the order of limits
is important and we find that
\begin{equation} \tilde{p}_{c=0} = \left\{
\begin{array}{ll}
0 & \textrm{for $\kappa < 4$}, \\
1 & \textrm{for $\kappa \geqslant 4$}.
\end{array}
\right.
\end{equation}
Note that in order to determine the phases, only the behavior of a
zero mode of $A$ at zero and at infinity is necessary. Examples of
traces of standard SLE are shown in Figures \ref{pick2} and
\ref{pick6}. Touching is visible for the $\kappa=6$ and apparently
does not occur for $\kappa=2$.

\subsection{Dynkin's formula with L\'evy flights}

The case of the generalized SLE is different from the Brownian
case: now $h(z,t)$ has discontinuities. Therefore, at the exit
time $T$ the value of $h(z,T)$ does not have to be $a$ or $b$: the
process can jump and overshoot the boundary of the interval. To
avoid this difficulty, we define $T$ to be the exit time from the
set
\begin{equation}
S = [-b,a) \cup (a,b],
\end{equation}
and also restrict ourselves to zero modes that are even functions
of $x$.

The process $h(z,t)$ can exit $S$ by either hitting $a$ (no
overshooting possible), or by taking a value beyond $-b$ or $b$,
in which case the overshooting is possible. But we consider the
limit $b \to \infty$. The behavior of zero modes, considered
below, is simple at infinity: they either diverge or are constant.
With these changes complete, we return to Eqs. (\ref{40},
\ref{50}).

The generator for the process $h(z,t)$ defined by Eqs. (\ref{11h},
\ref{SLEUL}) is (see \cite{apple}) $A = A_B + A_L$, where $A_B$ is
the same as in Eq. (\ref{A_B}), and
\begin{equation}
A_L f(x) = - \frac{c}{2\Gamma(-\alpha)
\cos\frac{\pi\alpha}{2}}\dashint(f(x + y) - f(x))\frac{dy}{|y|^{1
+ \alpha}}. \label{101}
\end{equation}
This operator contains the principal value integral, which ensures
convergence at small $y$ for $\alpha < 2$. The meaning of this
term is most transparent in the Fourier space, where it is
equivalent to the multiplication by $-c |k|^\alpha$.

Once again, we need to find a zero mode $f(x)$, obeying $Af = 0$.
These zero modes are studied in detail in the Appendix. Here we
summarize the results.

First of all, we find by direct calculation that the function
\begin{equation}
f_L(x) = \left\{ \begin{array}{ll} |x|^{\alpha - 1}, & \alpha \neq 1, \\
\ln |x|, & \alpha = 1
\end{array}
\right. \label{f_L}
\end{equation}
is a zero mode of $A_L$. Comparing this with Eq. (\ref{f_B}), we
conclude that $f_B(x) = f_L(x)$ if $1 - 4/\kappa = \alpha - 1$, or
\begin{equation}
\alpha = 2 - \frac{4}{\kappa}. \label{line}
\end{equation}
Along this line on the phase diagram we have an explicit
expression for a zero mode of the full operator $A$.

Secondly, for arbitrary $\alpha$ and $\kappa$ we can transform the
zero mode equation,  $Af=0$,  into  Fourier space:
\begin{equation}
\frac{d}{dk}\Big[\big(c|k|^\alpha  + \frac{\kappa}{2}k^2
\big)\tilde{f}(k)\Big] - 2k\tilde{f}(k) = 0,
\end{equation}
where $\tilde{f}(k)$ is the Fourier transform of $f(x).$ A
solution of this equation is
\begin{equation} \label{nuinu}
\tilde{f}(k) = |k|^{-\alpha} \Big(1  + \frac{\kappa}{2c} |k|^{2 -
\alpha} \Big)^{\frac{4}{(2 - \alpha)\kappa} - 1}.
\end{equation}
Apart from a multiplicative constant and a possible delta function
at $k=0$, this solution is unique. Its asymptotic behavior is
\begin{equation}  \label{1003}
\tilde{f}(k) \sim \left\{
\begin{array}{ll}
|k|^{\frac{4}{\kappa} - 2}, & k \to \infty, \\
|k|^{-\alpha}, & k \to  0.
\end{array}
\right.
\end{equation}
Note that Brownian motion determines the behavior at large $k$
(small distances in $x$-space), whereas L\'evy flights determine
the behavior at small $k$ (large distances in $x$-space).

Analysis of the asymptotic behavior of the inverse Fourier
transform of Eq. (\ref{nuinu}) gives the following:
\begin{equation} \label{1004}
f(x) \sim \left\{
\begin{array}{ll}
|x|^{1 - \frac{4}{\kappa}} & \textrm{for $\kappa\neq 4$ and
$x \to  0$}, \\
\log|x| & \textrm{for $\kappa=4$ and $x \to  0$}, \\
|x|^{\alpha - 1} & \textrm{for $\alpha\neq 1$ and
$x \to \infty$}, \\
\log|x| & \textrm{for $\alpha = 1$ and $x \to \infty$}.
\end{array}
\right.
\end{equation}
Strictly speaking, we are able to obtain these asymptotics only
for $\kappa > 2$. The trouble is that the function that behaves as
$|x|^{1 - 4/\kappa}$ with $\kappa < 2$ for small $x$ would lead to
a divergent integral in Eq. (\ref{101}), and so cannot be a zero
mode of $A$. Nevertheless, on physical grounds we believe that
there is nothing special happening at $\kappa = 2$, and that our
results for phases that we state next, apply to all values $\kappa
> 0$.

\begin{figure}[t]
\centering
\includegraphics[scale=0.5]{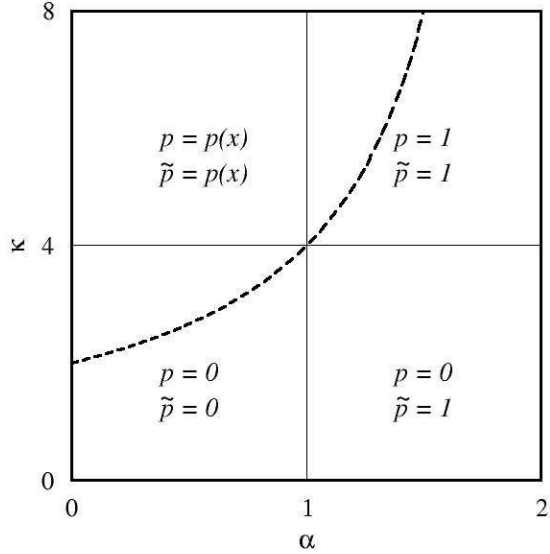}
\caption{The phase diagram of the generalized SLE (\ref{12}). The
dashed line is the hyperbola given by Eq. (\ref{line}) in the
text. Along this line an explicit expression for the zero mode of
the generator $A$ is known, see Eq. (\ref{f_L}).}
\label{pic-PhaseDiagram}
\end{figure}

The use of the asymtotic behavior (\ref{1004}) in the formulas
(\ref{40}, \ref{50}) gives the phases of SLE driven by the noise
(\ref{SLEUL}):
\begin{itemize}

\item

\textbf{isolated trees} $\kappa \leqslant 4$, $\alpha < 1$:  $p =
0$, $\tilde{p} = 0$. An example of the trace is shown in Figure
\ref{pick2m07}. The L\'evy flights have broad distribution. Long
jumps occur often. The trace can jump far from the existing trees
and then a new isolated tree starts growing. The chances of
returning to one of the old trees are small, so the old trees stop
growing. We hope to make this point quantitatively in a future
publication.

\item

\textbf{a forest of trees} $\kappa \leqslant 4$, $1 \leqslant
\alpha <2$: $p = 0$, $\tilde{p} = 1$. An example of the trace is
shown in Figure \ref{pick2m13}. The L\'evy flights rarely produce
very big jumps, so that the trace starts a new tree close to the
old one, or does not leave the old tree at all. The hull looks
like trees growing close to each other with their branches almost
touching.

\item

\textbf{a forest of bushes} $\kappa > 4$, $1 \leqslant \alpha <
2$: $p = 1$, $\tilde{p} = 1$. An example of the trace is shown in
Figure \ref{pick6m13}. The Brownian motion, which controls the
small scales, is stronger now. The trees become thicker and
different branches touch each other.

\item

\textbf{isolated bushes} $\kappa > 4$, $\alpha < 1$: $p$ and
$\tilde p$ are equal and finite, that is, the probability for a
fixed point $x$ on the real axis to be swallowed by the hull at a
finite time is a function $p(x)$ of the position of this point.
This function is defined by any of the equations (\ref{40},
\ref{50}) (the order of limits is unimportant in this phase),
where $f(x)$ is the inverse Fourier transform of (\ref{nuinu}).
$p(x)$ is a symmetric function decaying from $p(0) = 1$ to
$p(\infty) = 0$ with the asymptotic behavior:
\begin{equation}
p(x) \sim |x|^{\alpha - 1} \quad \textrm{for } x \to \infty, \quad
1 - p(x) \sim |x|^{1 - \frac{4}{\kappa}} \quad \textrm{for } x \to
0.\label{asymp}
\end{equation}
An example of the trace is shown in Figure \ref{pick6m07}. Since
$p(x)$ varies between 0 and 1, parts of the real axis will not be
swallowed, leaving gaps among trees.

\end{itemize}

As long as $c>0$, the first three phases do not depend on $c$ at
all. This is an expected property, because unlike $\kappa$ and
$\alpha$, $c$ flows under the renormalization (\ref{resc}) and $p$
and $\tilde{p}$ are constants. A phase diagram of the generalized
SLE is shown in Fig. \ref{pic-PhaseDiagram}.

\subsection{Probability $p(x)$ of swallowing on the real axis}

\begin{figure}[t]
\centering
\includegraphics[scale=0.55]{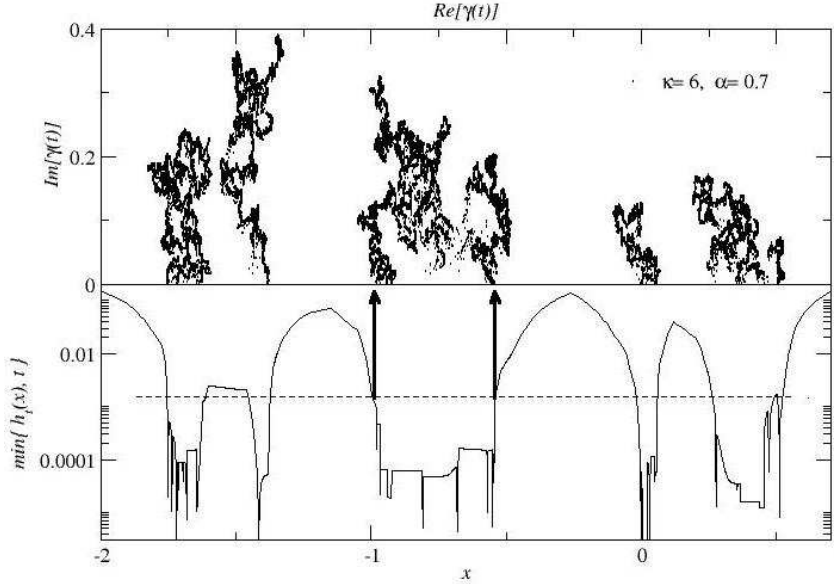}
\caption{We calculate which points on the real axis plane have
been swallowed by calculating $H(x)=min\{h(x, t); t<T\}$. There is
a direct correspondence between the areas swallowed by the trace
on the physical plane and the process $h(x, t)$ on the
mathematical plane. } \label{pic-onaxis-COR}
\end{figure}

Figures \ref{pick6}, \ref{pick6m07} and \ref{pick6m13} seem to
show that the trace swallows whole areas and touches itself.  In
the numerical approximation where the forcing consists of
discontinuous steps the trace doesn't exactly touch itself,
however it comes closer to touching as $\tau \to  0$ and whole
areas thus appear to be swallowed. We can then try to determine
the probabilities for  swallowing by setting a small-distance
cut-off and looking for limiting behavior as the cut-off goes to
zero.

In order to measure $p(x)$ we consider the real stochastic process
$h(x,t)$ on the mathematical plane defined by Eq.  (\ref{11h}). A
point $x$ on the axis is considered to be swallowed at time $T$ if
$|h(x, t)| < H_c$ for some $t < T$. Our method for determining
$H_c$ is shown in Figure \ref{pic-onaxis-COR}: For one particular
realization of the noise we calculate $H(x)=\min\{|h(x,t)|, t
\leqslant T\}$ for $x$ on the real axis. We use the same
realization to draw the trace on the physical plane. From the
trace we can infer which points of the axis have been surrounded
by the curve and thus have been swallowed. In the figure the
boundaries of a swallowed area on the real axis are marked by the
arrows; the dashed line corresponds to the cut off $H_c$, we see
that all points at which $H(x) < H_c$ appear to be swallowed in
the physical plane. We repeat for several noise realizations to
determine the value of the cut off accurately. The existence of
the cut-off means we will be unable to distinguish between $p$ and
$\tilde{p}$, so that the case $\alpha>1$, $\kappa<4$ (the forest
of trees) will be similar to $\alpha>1$, $\kappa>4$ (the forest of
bushes). For all cases we have taken $T \gg t^* =
\left(\displaystyle\frac{\kappa^\alpha}{c^2}\right)^{1/(2 -
\alpha)}$.

\begin{figure}[t]
\centering
\includegraphics[scale=0.55]{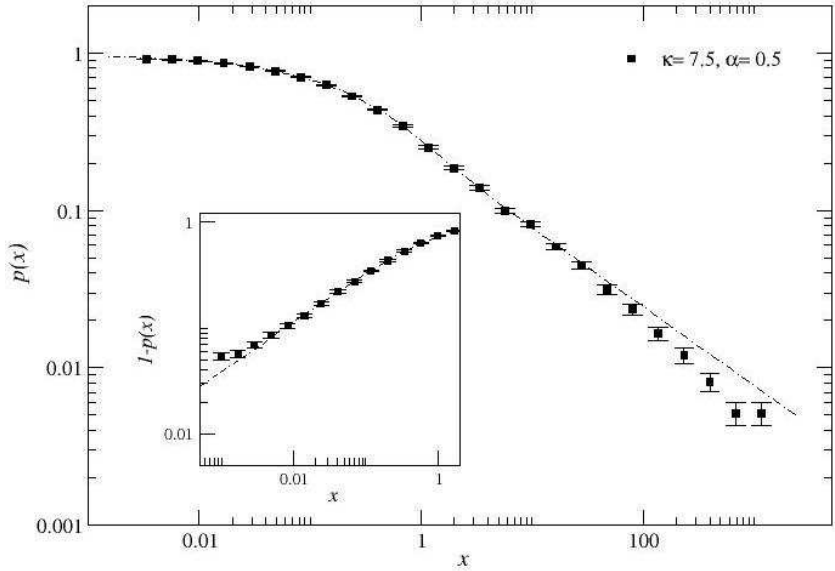}
\caption{SLE with L\'evy Flights and Brownian
component($\kappa=7.5, \alpha=0.5, c=10$). Main - Large $x$
behavior: Probability $p(x)$ of a point x on the real axis not to
be swallowed until time $T=10$ ($\tau=10^{-5}$, 1000000 steps,
$t^*=0.09$, $H_c=3~10^{-3}$, statistics for 7000 independent noise
realizations). Inset - Small $x$ behavior: Probability $1-p(x)$
($\tau=10^{-6}$, 1000000 steps, $H_c=6~10^{-4}$, statistics for
3000 independent noise realizations). In both cases the dotted
line is the theoretical prediction (Eq. (\ref{1004})) which is
followed accurately by the numerics for four decades of $x$.
Deviations for small $x<0.003$ are due to the finite time step
$\tau$ of the noise realizations; the data points approach the
theoretical prediction as we approach the scaling limit (not
shown)} \label{pic-onaxis-BL-1}
\end{figure}

Figure \ref{pic-onaxis-BL-1} shows the results of the probability
distribution $p(x)$ of a point on the real axis being swallowed by
the time $T = 10$ for a case with $\alpha < 1$, $\kappa > 4$
(isolated bushes). The numerical results closely follow the
theoretical prediction of Eq. (\ref{asymp}).

In general, we reproduce the theoretical form of $p(x)$ rather
well in this phase, but less well in the other phases. This is not
surprising, it is the effect of finite time $T$ in our
calculation. In the phases with forests, points are surrounded by
a process which includes multiple returns to the same region of
$x$.  In the simulations, not enough time is allowed for this
multiple return to work itself out. Furthermore, our use of the
cut-off means we are unable to distinguish between $p$ and
$\tilde{p}$, so that the method makes the forest of trees look
similar to the forest of bushes. Thus, we shall have to put
together a more refined method of analysis and do much longer runs
to make this distinction.

\section*{Discussion and conclusions}

The generalization of stochastic Loewner evolution, suggested in
this paper, has many new features. The  hull acquires branching
and the growth becomes a tree-like stochastic process. The phases
of such hull are very different from the usual SLE. Although the
driving force of SLE is still self-similar, the growing hull is
not, the force of the L\'evy flights increases with time, being a
relevant parameter under size rescaling.

The non-stationary nature of the growth considered in this paper
is related to a technical difficulty in its study. Many properties
of the standard SLE are obtained using the so-called random time
change which, in the case of the Brownian forcing, is consistent
with the form of the Loewner equation (\ref{11}). In the case of
L\'evy processes a similar random time change exists, but changes
the nature of the Loewner equation (\ref{12}) and, therefore,
cannot be used to extract useful information.

In terms of the geometry of the generalized SLE driven by L\'evy
processes, the non-stationarity means that many such properties
as, for example, the fractal dimension of the trace cannot be
globally defined. Instead, they may possibly be defined locally at
each stage of the evolution. The question of how one can
characterize the geometry of the generalized SLE will be addressed
in a future publication.

\section*{Acknowledgments}

This research was supported in part by NSF MRSEC Program under
DMR-0213745. IG is an Alfred P. Sloan Research Fellow, and was
also supported by an award from Research Corporation and the NSF
Career Award under DMR-0448820. We wish to acknowledge many
helpful discussions with Paul Wiegmann, Eldad Bettelheim, and
Seung Yeop Lee. IG also acknowledges useful communications with
John Cardy and Nigel Goldenfeld.

\appendix

\setcounter{section}{1}

\section*{Appendix}

In this Appendix we study zero modes of the operator $A = A_B +
A_L$, where
\begin{equation}
A_B = \frac{2}{x}\partial_x  + \frac{\kappa}{2}\partial_x^2, \quad
A_L f(x) = C \dashint \frac{f(y + x) - f(x)}{|y|^{1 + \alpha}} dy.
\end{equation}

\subsection*{Direct approach}

A zero mode of $A_B$ is given in Eq. (\ref{f_B}). To find zero
modes of $A_L$, let us consider the integral
\begin{equation}
I_{\alpha,\beta}(x) = \dashint \frac{|y + x|^\beta -
|x|^\beta}{|y|^{1 + \alpha}} dy. \label{Ialphabetax}
\end{equation}
This integral converges at large $y$ if $\beta < \alpha$, at small
$y$ for any $\beta$ and any $\alpha < 2$ (because of the principal
value), and at $y \approx -x$ for $\beta > -1$. Thus, we assume
$-1 < \beta < \alpha <2$.

By rescaling $y \to |x|y$ we reduce the integral
(\ref{Ialphabetax}) (for nonzero $x$) to
\begin{equation}
I_{\alpha,\beta}(x) = I_{\alpha,\beta} \, |x|^{\beta - \alpha},
\qquad I_{\alpha,\beta} = \dashint \frac{|y + 1|^\beta - 1}{|y|^{1
+ \alpha}} dy.
\end{equation}
The last integral here can be evaluated. First, we split the
integration interval into three intervals and change variables in
each appropriately:
\begin{eqnarray}
I_{\alpha,\beta}^\epsilon &=&  \Big(\int_{-\infty}^{-1} +
\int_{-1}^{-\epsilon} + \int_\epsilon^\infty \Big)\frac{|y +
1|^\beta - 1}{|y|^{1 + \alpha}} dy \nonumber \\
&=& \int_1^\infty \frac{(y - 1)^\beta - 1}{y^{1 + \alpha}} dy +
\int_\epsilon^1 \frac{(1 - y)^\beta - 1}{y^{1 + \alpha}} dy +
\int_\epsilon^\infty \frac{(y + 1)^\beta - 1}{y^{1 + \alpha}} dy.
\nonumber
\end{eqnarray}
The integrals here can be found in Ref. \cite{PBM1}. Some of them
contain hypergeometric functions that have to be transformed
further by applying the Kummer's connection formulas (see Sec.
2.9. in Chapter 2 of Ref. \cite{BE}). Combining all the results,
we obtain
\begin{eqnarray}
I_{\alpha,\beta}^\epsilon &=& \frac{\Gamma(\beta + 1)\Gamma(\alpha
- \beta)}{\Gamma(\alpha + 1)} + \frac{\Gamma(\beta +
1)\Gamma(-\alpha)}{\Gamma(\beta - \alpha + 1)} +
\frac{\Gamma(\alpha - \beta) \Gamma(-\alpha)}{\Gamma(-\beta)}
\nonumber \\
&& + \frac{1}{\alpha} \epsilon^{-\alpha} \big({}_2 F_1(-\alpha,
-\beta; 1 - \alpha; \epsilon) + {}_2 F_1(-\alpha, -\beta; 1 -
\alpha; -\epsilon) - 2 \big). \nonumber
\end{eqnarray}
Finally, in the limit $\epsilon \to 0$, the last term here
vanishes, and we get
\begin{equation}
I_{\alpha,\beta} = \frac{\Gamma(\beta + 1)\Gamma(\alpha -
\beta)}{\Gamma(\alpha + 1)} + \frac{\Gamma(\beta +
1)\Gamma(-\alpha)}{\Gamma(\beta - \alpha + 1)} +
\frac{\Gamma(\alpha - \beta) \Gamma(-\alpha)}{\Gamma(-\beta)}.
\label{Ialphabeta}
\end{equation}
Note that this expression vanishes identically if $\beta = \alpha
- 1$, thus showing that $f_L(x) = |x|^{\alpha - 1}$ is a zero mode
of $A_L$.

Let us now consider the limit of $\beta \to 0$. Specifically, we
note that
\begin{eqnarray}
J_\alpha(x) &=& \lim_{\beta \to 0}\beta^{-1}I_{\alpha,\beta}(x) =
\dashint \frac{\ln\Big|\displaystyle\frac{y}{x} + 1\Big|}{|y|^{1 +
\alpha}} dy
= J_\alpha |x|^{-\alpha}, \nonumber \\
J_\alpha &=& \dashint \frac{\ln|y + 1|}{|y|^{1 + \alpha}} dy.
\end{eqnarray}
The last integral can be obtained form Eq. (\ref{Ialphabeta}) by
carefully taking the limit $\beta \to 0$. Expanding gamma
functions we get
\begin{equation}
\beta^{-1}I_{\alpha,\beta} = \frac{1}{\alpha}\big(\psi(1 - \alpha)
- \psi(\alpha) \big) - \Gamma(\alpha) \Gamma(-\alpha) + O(\beta).
\end{equation}
Finally, using properties of the gamma and digamma functions, we
get
\begin{equation}
J_\alpha = \frac{\pi}{\alpha} \cot \frac{\pi\alpha}{2}.
\end{equation}
This expression vanishes at $\alpha = 1$ showing that $f_L(x) =
\ln |x|$ is a zero mode of $A_L$ for $\alpha = 1$.

Thus, we have shown that
\begin{equation}
f_L(x) = \left\{ \begin{array}{ll} |x|^{\alpha - 1}, & \alpha \neq 1, \\
\ln |x|, & \alpha = 1
\end{array}
\right.
\end{equation}
is a zero mode of $A_L$.

Next we note that $f_B(x) = f_L(x)$ if $1 - 4/\kappa = \alpha - 1$
or $\alpha = 2 - 4/\kappa$. Along this line on the phase diagram
we have an explicit expression for a zero mode of the full
operator $A$, which behaves for small and large values of $x$
exactly as claimed in the text of the paper. However, this is
restricted to the range of $\kappa
> 2$ (because $\alpha$ must be positive). This is, actually, more
generally true. Indeed, the integral $I_{\alpha, \beta}(x)$ with
$\beta = 1 - 4/\kappa$ converges at $y$ near $-x$ only for $\kappa
> 2$. Thus, our claims in the paper about the phases can be justified
strictly speaking only for $\kappa > 2$.

\subsection*{Fourier transform}

After the Fourier transform the zero mode equation $Af=0$ becomes
a differential equation
\begin{equation}
\frac{d}{dk}\Big[\big(c|k|^\alpha  + \frac{\kappa}{2}k^2
\big)\tilde{f}(k)\Big] - 2k\tilde{f}(k) = 0,
\end{equation}
Let us consider $k>0$. Introducing the new function
\begin{equation}
g(k) = \Big(c k^\alpha  + \frac{\kappa}{2}k^2 \Big)\tilde{f}(k),
\end{equation}
we rewrite the differential equation as
\begin{eqnarray}
\frac{dg}{g} &=& \frac{2k}{c k^\alpha  +
\displaystyle\frac{\kappa}{2}k^2} \, dk = \frac{2k^{1 - \alpha}}{c
+ \displaystyle\frac{\kappa}{2}k^{2 - \alpha}}
\, dk \nonumber \\
&=& \frac{4}{(2 - \alpha)\kappa} d \ln \Big(c  +
\displaystyle\frac{\kappa}{2}k^{2 - \alpha}\Big).
\end{eqnarray}
This is immediately integrated to give
\begin{equation}
g(k) = \Big(c  + \displaystyle\frac{\kappa}{2}k^{2 -
\alpha}\Big)^{\frac{4}{(2 - \alpha)\kappa}}.
\end{equation}
The case $k < 0$ is treated similarly, and we get in the end an
even function (as expected):
\begin{equation}
\tilde{f}(k) = |k|^{-\alpha} \Big(c  +
\displaystyle\frac{\kappa}{2}|k|^{2 - \alpha}\Big)^{\frac{4}{(2 -
\alpha)\kappa} - 1}.
\end{equation}
Apart from a multiplicative constant and a possible delta function
at $k=0$, this solution is unique.

Performing the inverse Fourier transform, we obtain a
representation of our zero mode:
\begin{equation}
f(x) = \int_{-\infty}^\infty \frac{dk}{2\pi} e^{ikx} \tilde{f}(k)
= \frac{1}{\pi} \int_0^\infty dk \cos(kx) \tilde{f}(k).
\label{IFT1}
\end{equation}
This integral diverges at the lower limit for $\alpha > 1$, but
this is a spurious divergence that is easy to amend. Namely, we
can formally subtract a (possibly infinite) constant from $f(x)$
replacing it by
\begin{equation}
f(x) = \frac{1}{\pi} \int_0^\infty dk (\cos(kx) - 1) \tilde{f}(k)
= \frac{2}{\pi} \int_0^\infty dk \sin^2 \Big(\frac{kx}{2}\Big)
\tilde{f}(k). \label{subtraction}
\end{equation}
This integral converges at the lower limit for all values $\alpha
< 2$, which is all we need.

Situation with convergence at the upper limit is trickier. By
Chartier's criterion (see Ref. \cite{WW}), the integral
(\ref{IFT1}) converges at the upper limit only if $\kappa
> 2$. This is the same condition that was obtained in the end of
the previous subsection where the Fourier transform was not used
at all. So, this seems to be a more problematic issue. We also
note that after the subtraction performed in Eq.
(\ref{subtraction}), the convergence at the upper limit worsens.
Now the integral is convergent only for $\kappa > 4$. But this can
be ignored. We simply perform the integration for $\kappa > 4$ and
then analytically continue to smaller values of $\kappa$.
Convergence also improves when we take derivative in either of
Eqs. (\ref{IFT1}, \ref{subtraction}) with respect to the strength
of the L\'evy term $c$. We can then in principle obtain $f(x)$
integrating in $c$.

Another comment is in order here. If we tried to obtain the zero
mode $f_B(x)$ for the Brownian SLE case using the Fourier
transform, we would get
\begin{equation}
f(x) = \int_{-\infty}^\infty \frac{dk}{2\pi} e^{ikx}
|k|^{\frac{4}{\kappa} - 2} = \frac{1}{\pi} \int_0^\infty dk
\cos(kx) \, k^{\frac{4}{\kappa} - 2} \label{IFT2}
\end{equation}
instead of Eq. (\ref{IFT1}). The last integral has the same
behavior at infinity as the one in Eq. (\ref{IFT1}). But the zero
mode $f_B(x)$ of $A_B$ given in Eq. (\ref{f_B}) exist for all
values of $\kappa$. This is another reason for us to believe that
the value $\kappa = 2$ is not special, and that our results for
the phases of the generalized SLE apply for all values of
$\kappa$.

Consider now the integral
\begin{equation}
f(x,\kappa,\alpha) = \int_0^\infty dk \, k^{-\alpha} \Big(1  +
k^{2 - \alpha}\Big)^{\frac{4}{(2 - \alpha)\kappa} - 1} \sin^2
\Big(\frac{kx}{2}\Big),
\end{equation}
where we ignored the overall constant coming from pulling a power
$c$ out of the integral, as well as rescaling of $k$ and $x$. This
integral can be evaluated explicitly through hypergeometric
functions for such values of $\alpha$ that
\begin{equation}
2 - \alpha = \frac{2p}{q}, \label{alpha}
\end{equation}
where $p, q \in \mathbb{N}$ are mutually prime, see Ref.
\cite{PBM1}, Eq. 2.5.2.3. The Mathematica software can also do the
integral with these values of $\alpha$. The simplest of these
corresponds to $p=1, q=2$, that is, $\alpha = 1$. The
corresponding integral is
\begin{eqnarray}
f(x,\kappa,1) &=& \int_0^\infty dk \, k^{-1} (1 +
k)^{\frac{4}{\kappa} - 1} \sin^2
\Big(\frac{kx}{2}\Big) \nonumber \\
&=& \frac{\pi  \sec \left(\frac{2 \pi }{\kappa }\right)}{4 \Gamma
\left(2 - \frac{4}{\kappa }\right)}   \,
_1F_2\left(\frac{1}{2}-\frac{2}{\kappa
};\frac{1}{2},\frac{3}{2}-\frac{2}{\kappa };-\frac{x^2}{4}\right)
 x^{1-\frac{4}{\kappa }}
\nonumber\\
&& -\frac{\pi  \kappa  \csc \left(\frac{2 \pi }{\kappa }\right)}{8
(\kappa -2) \Gamma \left(\frac{\kappa -4}{\kappa }\right)}  \,
_1F_2\left(1-\frac{2}{\kappa };\frac{3}{2},2-\frac{2}{\kappa
};-\frac{x^2}{4}\right) x^{2 - \frac{4}{\kappa}}
\nonumber\\
&& + \frac{\kappa^2}{16(\kappa + 4)} \,
_2F_3\left(1,1;2,1+\frac{2}{\kappa },\frac{3}{2}+\frac{2}{\kappa
};-\frac{x^2}{4}\right) x^2.
\end{eqnarray}

For all values of $\alpha$ given by Eq. (\ref{alpha}) the
structure of the answers is similar. They are given in terms of
hypergeometric functions $\, _p F_q$. The asymptotic analysis of
these expressions can be performed with the help of Ref.
\cite{Luke}, and in all cases gives the result given in Eq.
(\ref{1004}).

\end{document}